\documentclass[letterpaper, 10 pt, conference]{ieeeconf}

\usepackage{soul}
\usepackage{fullpage}
\usepackage{epsfig,graphicx,float}
\usepackage{xcolor}
\usepackage{verbatim}
\usepackage{subfigure}
\usepackage{url}
\usepackage{amsmath,amstext,amssymb,txfonts,bm}
\usepackage{amsfonts}
\usepackage{tabularx}
\usepackage{times}
\usepackage{multirow}
\usepackage{url}
\usepackage{cite}
\usepackage{setspace}
\usepackage{framed}
\usepackage{mathrsfs}
\usepackage{sansmath}
\usepackage{mathtools}
\usepackage{wrapfig}
\usepackage{caption}
\usepackage{graphicx}
\usepackage{mwe}
\usepackage{subfig}
\usepackage{cancel}
\usepackage{geometry}
\IEEEoverridecommandlockouts

\title{\bf Control-Oriented Models Inform Synthetic Biology Strategies in CAR T Cell Immunotherapy}
\author{Raffaele Romagnoli
\thanks{R. Romagnoli is with the School of Science and Engineering, Duquesne University, 5000 Forbes Ave, Pittsburgh, PA 15213, USA:
        {\tt\small romagnolir@duq.edu}}%
}

\begin{document}
\maketitle
\begin{abstract}
    Chimeric antigen receptor (CAR) T cell therapy is revolutionizing the treatment of blood cancers. Mathematical models that can predict the effectiveness of immunotherapies such as CAR T are of increasing interest due to their ability to reduce the number of experiments performed and to guide the theoretical development of new therapeutic strategies. {Following this rationale, we propose the use of control-oriented models to guide the augmentation of CAR T therapy with synthetic gene circuitry. Here we present an initial investigation where we adapt a previously developed CAR T model for control-oriented purposes. We then explore the impact of realistic alternative activation methods as control inputs to ensure effective tumor clearance.}
    
\end{abstract}
\section{Introduction}

Chimeric antigen receptor (CAR) T cells have shown remarkable therapeutic potential in the
treatment of hematological cancers such as B cell leukemias \cite{park2018long} and lymphomas \cite{june2018chimeric}, \cite{roschewski2022car}. CAR T cells are
genetically engineered to interact with high specificity to tumor antigens \cite{gross1989expression}. Binding of the CAR to tumor target antigens leads to activation and proliferation of CAR T cells and tumor cell lysis \cite{june2018chimeric}. Yet, CAR T therapies have encountered an array of different issues such as insufficient numbers of CAR T
cells, immunosuppressive signaling, \cite{chang2018rewiring}, and
outgrowth of target antigen-negative or low tumor cells \cite{cho2018universal,zah2016t}. A promising strategy to overcome these challenges is through augmentation of CAR T cells with additional synthetic gene circuitry \cite{labanieh2023car, roybal2016precision, chmielewski2015trucks}. Current synthetic biology control strategies are focused on achieving more specific activation of CAR T cells in presence of the tumor, as well as controlling therapeutic timing, location, and strength. \cite{brandt2020emerging}. 

The concept of control in CAR T immunotherapy is still removed from the classical control theory, the incorporation of which may provide important insights and tools to design more effective therapies, and this manuscript aims to bridge this theoretical gap. Two major classes of models have been previously developed to investigate CAR T cell behaviors: agent-based models (ABMs) and models using partial differential equations/ordinary differential equations (PDEs/ODEs). ABMs are computational methods that simulate the spatial and temporal behavior of every single agent/cell that interacts with the environment following specific rules \cite{prybutok2022mapping}. This kind of model is based on simulations to capture a specific behavior of the system which can be difficult to describe with mathematical tools \cite{an2017optimization}. In contrast mathematical models based on PDEs and ODEs allow analyses that are impossible to make with ABMs or solve problems such as optimization and control. ODEs have been used in the framework of CAR T cells to describe the behavior of single cells \cite{qi2022cellular}, however, ODEs can also be applied to additional mechanistic levels in the study of CAR T cells. For instance, a specific CAR-T design at the cellular level will have effects on the entire system/population of CAR T cells and will act on a population of tumor cells. A previously developed model that provides a high-level behavior of tumor cell and CAR T cell populations for non-solid tumors is called CART\textit{math} \cite{barros2021cart}. This model introduces three ODEs that describe dynamics and the interactions between the population of active CAR T cells, memory (non-active) CAR T cells, and tumor cells. This model mimics a predator-prey-like model where from the theoretical viewpoint tumor clearance cannot be achieved, which conflicts with real-world outcomes. Additionally, this model correctly demonstrates that  the therapeutic success is highly dependent on input CAR T cell number. Since this model reduces all the effects of possible synthetic biology strategies in the design of the CAR T cells into a set of parameters, it is difficult to make a direct connection between the effectiveness of the therapy and potential synthetic gene circuitry designs. {In this paper, we tackle the challenge of achieving tumor clearance from a control perspective. We demonstrate the potential effectiveness of an alternative CAR T cell activation strategy by modeling it as an external control input. This additional input complements the traditional activation mechanism driven by the presence of tumor cells. Our approach builds upon the model introduced in \cite{barros2021cart}. We propose a novel model and a backstepping analysis \cite{Khalil:1173048} to describe the effects of this control input on tumor clearance in terms of asymptotic stability. Additionally, we describe novel activation strategies for CAR T cells that are aligned with our model.} 

 \section{CAR T Cell Description} \label{sec:case_study}
 CAR T cells contain synthetic receptors created by fusing multiple proteins together in a specific
orientation. Generally, CARs consist of an antigen binding domain, usually a single chain
variable fragment (scFv) and T cell signaling domains such as CD28, 4-1BB, and CD3$\zeta$\cite{labanieh2023car}. The therapy is
manufactured typically by removing a patient’s T cells, genetically engineering the cells, and infusing them back into
the patient\cite{brudno2018chimeric} The cells act as a living therapeutic capable of specifically targeting the patient’s immune system to
fight cancer. Binding of CAR T cells to tumor antigens causes T cell activation through the
signaling domains with endogenous downstream mechanisms, functionally rewiring T cell
activation to user-defined antigen recognition\cite{june2018chimeric} .
\begin{figure}[!ht]
    \centering    \includegraphics[width=1\columnwidth]{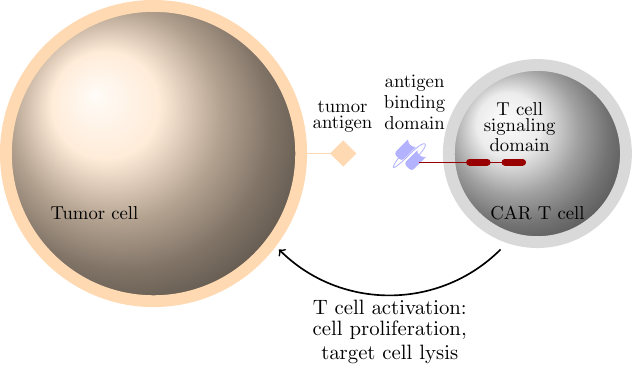}
    \caption{Schematic of a CAR T cell recognizing a tumor cell.}
    \label{fig:cart_scheme}
\end{figure}

\section{Ordinary Differential Equation (ODE) model}
CAR T cells are genetically modified to recognize and eliminate tumor cells, and a mathematical model to predict their therapeutic activity has been developed in \cite{barros2021cart}. This model provides ordinary differential equations based on the scheme of Fig. \ref{fig:cart_scheme}.
\begin{figure}[!ht]
    \centering    \includegraphics[width=1\columnwidth]{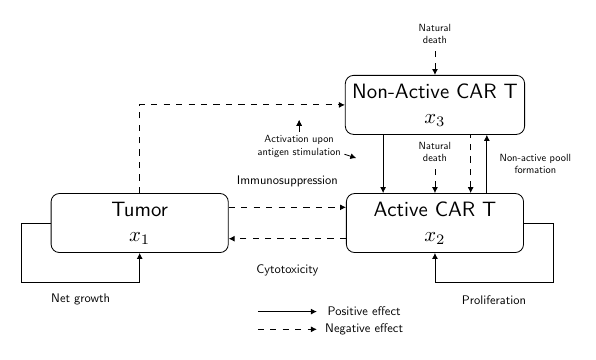}
    \caption{Interaction scheme between non-active CAR T cells $x_3$, active CAR T cells $x_2$, and tumor cells $x_1$. This scheme is adapted from \cite{barros2021cart}.}
    \label{fig:cart_scheme}
\end{figure}
The active CAR T cells have a cytotoxic effect on tumor cells. These cells proliferate, differentiate into non-active CAR T cells, and die naturally or are impaired by tumor-induced immunosuppressive mechanisms. Non-active  CAR T cells also die naturally, and they activate when stimulated by the tumor-associated antigen turning into active CAR T cells. Finally, the tumor cells grow based on the available resources and are killed by effector CAR T cells. In Fig. \ref{fig:cart_scheme} the solid and dashed arrows represent  positive and negative effects due to interactions between the three different kinds of cells. For example, the non-active pool formation represents the situation where active CAR T cells become non-active, this phenomenon has a positive effect on non-active CAR T and a negative effect on active CAR T cells. Finally, the variables $x_1$, $x_2$, and $x_3$ represent the tumor cells, the active and non-active CAR T cells respectively. Since these variables represent the number of cells, they are all positive. Moreover, all the model parameters that will be defined in the next subsections are also non-negative.

\subsection{ Tumor cells}
The tumor cells dynamics are
\begin{equation}\label{eqn:tumor}
    \dot x_1=rx_1(1-bx_1)-\gamma x_2 x_1
\end{equation}
where 
\begin{itemize}
    \item $r$: maximum growth rate of tumor cells ([day$^{-1}$]).
    \item $b$: inverse of the tumor carrying capacity ([cells$^{-1}$]).
    \item $\gamma$: Cytotoxic coefficient induced by effector CAR T cells ([(cell $\cdot$ day)$^{-1}$]).
\end{itemize}

The dynamical equation \eqref{eqn:tumor} considers the tumor cells to be bounded in their growth, and the population is reduced by the active CAR T cells $\gamma x_2 x_1$.

\subsection{Active CAR T cells}

The dynamics of active CAR T cells are given by
\begin{equation}\label{eqn:effector}
 \dot x_2= (\phi-\rho)x_2+\theta x_3 x_1 - \alpha x_2 x_1
\end{equation}
where the parameters are
\begin{itemize}
    \item $\phi$:  proliferation rate of the active CAR T cells ([day$^{-1}$]).
    \item $\rho$:  reduction rate of the active CAR T cells due to the natural death and differentiation into non-active CAR T cells([day$^{-1}$]).
    \item $\theta$: conversion coefficient of non-active CAR T into active CAR T cells due to the interaction with the tumor cells ([(cell $\cdot$ day)$^{-1}$]).
    \item $\alpha$: Inhibition/expansion coefficient of active CAR T cells due to the interaction with the tumor cells ([(cell $\cdot$ day)$^{-1}$]).  
\end{itemize}
 Based on equation \eqref{eqn:effector}, the number of active CAR T cells decays to zero in the absence of tumor cells if $\phi<\rho$. The term $\theta x_3 x_1$ represents the activation rate of non-active CAR T cells due to the interaction with the tumor, and the term $\alpha x_2 x_1$ describes the effects of stimulatory and inhibitory signals on active CAR T cells modulated by the tumor.

\subsection{Non-active CAR T cells}
The dynamics of non-active CAR T cells are
\begin{equation}\label{eqn:memory}
\dot x_3 = \epsilon x_2 - \theta x_3 x_1 - \mu x_3
\end{equation}
where
\begin{itemize}
    \item $\epsilon$: effective conversion rate of active CAR T into non-active CAR T cells ([day$^{-1}$]).
    \item $\mu$: death rate of non-active CAR T cells (day$^{-1}$). 
\end{itemize}
 If $\phi>\rho-\epsilon$ the active CAR T cells proliferate and become non-active CAR T cells. In this case, the term $\theta x_3 x_1$, which also appears in \eqref{eqn:effector}, has a negative effect on the non-active CAR T cell population.

\subsection{Model Analysis}
The above system can be rewritten as 
\[
\dot x = f(x)
\]
where $x=[x_1,x_2,x_3]^T$ and
\[
f(x)=\left[\begin{array}{l}
   r x_1 (1-b x_1)- \gamma x_2 x_1\\
   (\phi-\rho) x_2 + \theta x_3 x_1 - \alpha x_2 x_1 \\
   - \mu x_3  + \epsilon x_2 - \theta x_3 x_1
   \end{array}\right].
\]
This model has four equilibrium points \cite{barros2021cart} that can be admissible or not based on the value of the model's parameters. We compute the Jacobian matrix as a function of the equilibrium point $x_e$:
\begin{equation}
    \left[\frac{\partial f}{\partial x} \right]=\left[\begin{array}{ccc}
        r(1-2bx_1) & \gamma x_2& 0\\
       \theta x_3 -\alpha x_2 & (\phi-\rho)x_2-\alpha x_1  & \theta x_1 \\
         -\theta x_3 & \epsilon & -\mu - \theta x_1             
    \end{array} \right]
\end{equation}
Considering $x_e^0=[0,\;0,\;0]^T$, the above matrix turns into 
\[
\left[\frac{\partial f}{\partial x}\right]_{x=x_e^0}=\left[\begin{array}{ccc}
       r & 0 & 0 \\
       0  & \phi-\rho & 0  \\
       0 & \epsilon  & -\mu 
    \end{array} \right]_{x=x_e^0}
\]

which has a positive eigenvalue, hence $x_e^0$ is structurally unstable. In fact, the possible perturbation in parameters does not affect the stability property of $x_e^0$ that remains unstable. The presence of tumor cells makes this equilibrium point unstable, and if there are no CAR T cells, it can grow until reaching a maximum bound. This point corresponds to a second equilibrium point given by $x_e^1=[1/b,\;0,\; 0]^T$. By computing the Jacobian matrix 
\[
\left[\frac{\partial f}{\partial x}\right]_{x=x_e^1}\left[\begin{array}{ccc}
       -r & 0 & 0\\
       0 & \phi-\rho- \alpha/b  & \theta/b  \\
       0 & \epsilon  & -\mu-\theta/b
    \end{array} \right]_{x=x_e^1},
\]
it is possible to observe that variation in the model's parameters can make $x_{e}^1$ asymptotically stable or unstable. The other two equilibrium points $x_{e}^2$ and $x_{e}^3$ are less intuitive w.r.t the first two, and they can be calculated by following several steps that are shown in \cite{barros2021cart}. Due to the presence of bifurcations, $x_{e}^2$ and $x_{e}^3$ may not be feasible, e.g. their components may have negative values, or can be stable or unstable. Since $x_{e}^0$ is always unstable, the effectiveness of CAR-T therapy can be achieved if $x_{e}^1$ is unstable and one of the other admissible equilibrium points is asymptotically stable. If in this situation $x_{e}^1$ is also asymptotically stable then there is a condition of bi-stability, and the initial conditions are crucial to avoid the system converging to $x_{e}^1$. The feasible equilibrium points $x_{e}^2$ and $x_{e}^3$ describe the situation where tumor and CAR T cells coexist, and a small quantity of tumor cells is needed to make the active CAR T cells proliferate and keep the number of tumor cells bounded.



\subsection{Problem Statement}

The model developed in \cite{barros2021cart} can provide useful information about the dosing strategies to obtain specific initial conditions of the CAR T cells. In fact, despite theory predicting the impossibility of tumor clearance, it can be achieved in practice if the system can start from specific initial conditions of the tumor and CAR T cells. The experiments referenced in \cite{barros2021cart} show that using specific initial conditions can lead to low values of tumor cells as predicted by the model, and this result represents the practical situation where the tumor is eliminated. {However, the therapy is highly sensitive to these initial conditions and many arbitrary initial conditions for tumor additional tumor growth and disease relapse.} 
{In this paper, we illustrate the beneficial impact of an alternative strategy for CAR T cell activation on tumor clearance. We introduce an enhanced model incorporating an additional control input for CAR T cell activation, which can be represented as a feedforward input. Through a backstepping approach analysis, we demonstrate that this supplementary input ensures reliable tumor clearance by achieving asymptotic stability at the point $x_1=0$.}

\section{Control Oriented Model For Non-Solid Tumors}
Here we take the previous model as a baseline and propose a new control-oriented model to determine the conditions where the tumor clearance condition $x_1=0$ is asymptotically stable and the number of active and non-active CAR T cells converge to bounded values $\bar x_2$ and $\bar x_3$ respectively. 



\subsection{Proposed Model}
\begin{figure}[!htp]
    \centering
    \includegraphics[width=1\columnwidth]{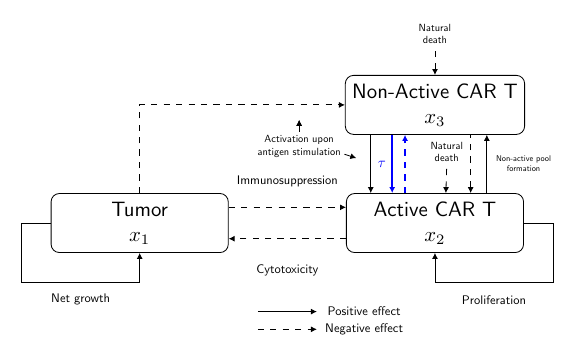}
    \caption{The modified model incorporates an additional input, denoted as $\tau$, which governs the activation of the CAR T cells.}
    \label{fig:solid}
\end{figure}

To develop a control-oriented model we first identify the controllable inputs of the system. From the dynamic equations  

\begin{equation} \label{eqn:solid_model}
    \left\lbrace \begin{array}{l}
      \dot x_1 = r x_1 (1-bx_1)-\gamma x_1 x_2   \\
      \dot x_2 = (\phi-\rho)x_2 - \alpha x_2 x_1 + \theta x_3 x_1 \\
      \dot x_3 = -\mu x_3 - \theta x_3 x_1 + \epsilon x_2 
    \end{array} \right.
\end{equation}
the system can be seen as the cascade connection between non-active and active CAR T cells and active CAR T and tumor cells. {In our analysis, we concentrate on the initial two equations of \eqref{eqn:solid_model}. Here, the non-active CAR T cells $x_3$ are regarded as an input $u$ influencing the dynamics of the active CAR T cells. Additionally, we introduce a distinct mechanism for activating the CAR T cells, which is represented by a positive supplementary input, denoted as $\tau$.} 
\begin{equation} \label{eqn:solid_model_sub1}
    \left\lbrace\begin{array}{l}
      \dot x_1 = r x_1 (1-bx_1)-\gamma x_1 x_2   \\
      \dot x_2 = (\phi-\rho)x_2 - \alpha x_2 x_1 + \theta x_1 u + \tau     
    \end{array} \right.
\end{equation}
Fig. \ref{fig:solid} shows the modified interaction scheme between tumor, active and non active CAR T cells. 

The control input $u$ represents the non-active CAR T cells to be activated by the presence of the tumor, and instead $\tau$ is a control input generated by an alternative mechanism of activation of the CAR T cells. The blue arrows signify positive and negative effects on the growth of the active CAR T cells, which also has a negative effect on the population of non-active CAR T cells.

\subsection{Asymptotic Stability using the Backstepping Approach}
The aim is to illustrate the beneficial impact of the additional input $\tau$ on tumor clearance. Specifically, focusing on the tumor dynamics outlined in the first equation of \eqref{eqn:solid_model_sub1}, we interpret the effects of the active CAR T cells $x_2$ as a control input $u_1$. To achieve tumor clearance, the control input $u_1$ must ensure asymptotic stability. Thus, we assume that $u_1$ acts as the following state feedback controller:
\begin{equation}
    u_1= \kappa(x_1)= \frac{r}{\gamma}\left(a-bx_1 \right) 
\end{equation}
which transforms dynamics of the tumor cells into
\[
\dot x_1 = r x_1 - rbx_1^2 - arx_1+rbx_1^2=r(1-a)x_1
\]
which is a linear ODE with $x_1=0$ asymptotically stable for any $a>1$. 
Our objective is to utilize backstepping control analysis to derive conditions under which the active CAR T cells exhibit behavior consistent with a specified state feedback controller $u_1$, aimed at driving the number of tumor cells to zero.
To do so, we backstep this variable by defining 
\begin{equation}
    z_2 \triangleq x_2 - \kappa(x_1) = x_2 - \frac{r}{\gamma}\biggl({a-bx_1} \biggl).
\end{equation}
The goal is to show that $z_2$ asymptotically converges to zero, hence $x_2$ converges to $\kappa(x_1)$. After a change of variables, the first equation of \eqref{eqn:solid_model_sub1} becomes 
\[
\dot x_1 = r(1-a)x_1 - \gamma x_1 z_2. 
\]
Instead the l.h.s of the second equation of \eqref{eqn:solid_model_sub1} is
\[
\dot x_2 = \dot z_2 - \hat b \dot x_1 = \dot z_2 - \hat b\biggl({r(1-a)x_1 - \gamma x_1 z_2}\biggl)
\]
where $\hat b = rb/\gamma$. The r.h.s is
\[
\biggl({\phi-\rho - \alpha x_1}\biggl)\left(z_2+\frac{r}{\gamma}(a-bx_1)\right)+\theta x_1 u+ \tau.
\]
Hence, 
\begin{eqnarray}
\dot z_2 &=& \biggl({\phi-\rho - \alpha x_1}\biggl)\left(z_2+\frac{r}{\gamma}(a-bx_1)\right)+ \hat br(1-a)x_1 \nonumber \\
&& - \hat b\gamma x_1 z_2 + \theta x_1 u + \tau \nonumber \\
&=& (\phi-\rho)z_2 + (\phi-\rho)\frac{r}{\gamma}a + \tau + \biggr[{\hat br(1-a) - \hat b\gamma z_2} \nonumber\\
&& \left. +\theta u - \hat b (\phi-\rho)- \alpha z_2 +\frac{\alpha\; r}{\gamma} \left(bx_1 - a \right) \right]x_1 
\end{eqnarray}

{We examine the following Lyapunov function candidate, as expressed in Equation \eqref{Lyapunov_f}:
\begin{equation}\label{Lyapunov_f}
    V(x_1,z_2)=\frac{1}{2}\left(\xi x_1^2 + z_2^2\right),
\end{equation}
where $\xi>0$ represents a positive real scalar value. The time derivative of \eqref{Lyapunov_f} is given by \eqref{v_dot} reported in Fig. \eqref{fig:dotV}.
\begin{figure*}
\begin{eqnarray}\label{v_dot}
\dot V (x_1,z_2) &=& \xi x_1 \dot x_1 + z_2 \dot z_2 \nonumber \\
&=& \xi r(1-a)x_1^2  + (\phi-\rho) z_2^2 + \left((\phi-\rho)\frac{r}{\gamma}a+\tau \right)z_2 \nonumber \\
&+& \left[\hat b r(1-a)-\hat b \gamma z_2+\theta u - \hat b(\phi-\rho)-\alpha z_2 + \alpha \frac{r}{\gamma}(bx_1-a)- \gamma \xi x_1\right]x_1 z_2
\end{eqnarray}
\caption{Derivation of $\dot V (x_1,z_2)$.}
\label{fig:dotV}
\end{figure*}
We assume now that 
\begin{equation} \label{tau}
    \tau=-(\phi-\rho)\frac{r}{\gamma}a,
\end{equation}
which cancels out the term $(\phi-\rho)(r/\gamma)az_2$ in $\dot V(x_1,x_2)$.
Consequently, by defining two positive scalars as $\hat \ell \triangleq \vert r(1-a)\vert$ and $\hat m \triangleq \vert\phi-\rho \vert$, we can make the following assumption in a neighborhood of $x_1=z_2=0$ on the term of \eqref{v_dot} that multiplies $x_1z_2$:
\begin{eqnarray} \label{bound_k}
   &&-\hat b \hat \ell-\hat b \gamma z_2+\theta u - \gamma \xi x_1+ \hat b\hat m-\alpha z_2 \nonumber \\
   && \hspace{2cm}+ \alpha \frac{r}{\gamma}(bx_1-a)-\gamma \xi x_1\leq 2k    
\end{eqnarray}
taking into account that $a>1$, $\phi-\rho<0$, and for the moment that $u$ is bounded. Therefore,  we can bound $\dot V(x_1,z_2)$ as follows:
\begin{eqnarray}
    \dot V(x_1,z_2) &\leq& - \xi \hat \ell x_1^2 - \hat m z_2^2 + 2k\vert x_1 \vert \vert z_2 \vert \nonumber\\
    &=&-\left[\begin{array}{c}
         \vert x_1 \vert  \\
         \vert z_2 \vert  
    \end{array}\right]^T \left[\begin{array}{cc}
        \xi \hat \ell & -k  \\
         -k & \hat m 
    \end{array}\right] \left[\begin{array}{c}
         \vert x_1 \vert  \\
         \vert z_2 \vert  
    \end{array}\right]
\end{eqnarray}
For this quadratic form to be positive definite, we require \cite{Khalil:1173048}:
\[
\xi \hat \ell \hat m - k^2 >0 \rightarrow k< \sqrt{\xi \hat \ell \hat m}
\]
An appropriate value of $\xi$ can be determined based on the bound $k > 0$. Consequently, \eqref{Lyapunov_f} becomes a valid Lyapunov function, establishing $x_1=z_2=0$ as an asymptotically stable equilibrium point.
The input $u$ corresponds to the variable $x_3$, representing the non-active CAR T cells. Consider the CAR T subsystem
\begin{equation}\label{lin_sys}
    \left.\begin{array}{l}
        \dot x_2 = (\phi-\rho)x_2 - \alpha x_2 x_1 + \theta x_3 x_1 + \tau \\
      \dot x_3 = -\mu x_3 - \theta x_3 x_1 + \epsilon x_2 -\tau
    \end{array} \right.
\end{equation}
with $\tau=0$ and $x_1=0$ the equilibrium point $x_2=x_3=0$ is asymptotically stable \cite{barros2021cart} hence if $\tau$ is bounded then the behavior of $x_2$ and $x_3$ remains bounded. From \eqref{lin_sys}, the equilibrium point for $x_1=0$ is given by 
\begin{equation}\label{new_eq}
    x_n^e = \left[\begin{array}{ccc}
       0 & \bar x_2  & \bar x_3 \\
    \end{array} \right]^T,
\end{equation}
with
\[
\bar x_2=-\frac{\tau}{(\phi-\rho)};\quad \bar x_3= -\frac{\tau(\rho-\phi-\epsilon)}{(\phi-\rho)\mu},
\]
where $\bar x_2, \bar x_3$ are positive real numbers. Examining the first equation of \eqref{eqn:solid_model_sub1}, it becomes apparent that there exists a value of $x_1$ and a constant value of $x_2$ that render the equation asymptotically stable. Consequently, if $x_1$ is decreasing, then \eqref{lin_sys} exhibits bounded dynamics, ensuring the existence of a bound $k > 0$ for \eqref{bound_k}.
In summary, to establish $x_1=0$ as an attractive point while ensuring the boundedness of $x_2$ and $x_3$ for a given initial condition of tumor cells $x_1(0) \neq 0$, it is imperative to evaluate the initial conditions of the non-active CAR T cells $x_3(0)$. This guides the system's behavior in a vicinity of $x_1=0$, which serves as a point of attraction due to the external input $\tau$ calculated in \eqref{tau}.
Choosing $\tau$ to nullify a term in \eqref{v_dot} may prove challenging in practical implementation. However, the backstepping analysis relies on a generic $a>1$, ensuring asymptotic stability for the variable $x_1$ for any $\tau > \frac{(\rho-\phi) \cdot r}{\gamma}$. 
}

\subsection{Connection with existing works}
The new model introduced in Figure \ref{fig:solid} incorporates an additional activation mechanism $\tau$ \eqref{tau}, and the entire analysis is developed around it. The question arises whether $\tau$ can be implemented using synthetic gene circuitry. Recent studies by Chang et al. \cite{chang2018rewiring} and Allen et al. \cite{allen2022synthetic} align with our research, as they explore innovative methods to enhance CAR-T cell functionality beyond traditional antigen recognition.

Chang et al. \cite{chang2018rewiring} demonstrate the potential of utilizing CARs to recognize soluble ligands, a distinct class of biomarkers, thereby expanding their utility in disease treatment. Their study illustrates the engineering of CAR T cells to robustly respond to various soluble ligands, including the CD19 ectodomain, GFP variants, and transforming growth factor beta (TGF-b). They reveal that CAR signaling triggered by soluble ligands depends on ligand-mediated CAR dimerization and can be finely regulated by adjusting the mechanical coupling between the CAR's ligand-binding and signaling domains. These findings suggest a role for mechanotransduction in CAR signaling and propose a systematic approach to engineering immune-cell responses to soluble extracellular ligands.

In contrast, Allen et al. \cite{allen2022synthetic} discuss the challenges facing chimeric antigen receptor (CAR) T cells in combating solid tumors within immunosuppressive microenvironments. To address this, they engineer circuits utilizing tumor-specific synNotch receptors to locally induce production of the cytokine IL-2. These circuits significantly enhance CAR T cell infiltration and clearance of immune-excluded tumors, without causing systemic toxicity. The most effective IL-2 induction circuit operates in an autocrine and T cell receptor (TCR) or CAR-independent manner, effectively bypassing suppression mechanisms such as IL-2 consumption or inhibition of TCR signaling. Their work demonstrates the feasibility of reconstructing synthetic T cell circuits capable of activating the necessary outputs for an antitumor response while evading critical points of tumor suppression.

{\section{Simulation Results}
For simulations, we consider the parameters reported in Table 2, HDLM-2 + CAR-T 123 of \cite{barros2021cart} and the experiments reported in Section 3.1 a, where at day 42 CAR T cells are injected.  In Fig. \ref{fig:no_control} we report the behavior of the model \cite{barros2021cart} with initial conditions $x(0)=[0,\; 0,\; 2\times 10^6]$. At day 42, a population of inactive CAR T cells $x_3=0.6\times 10^6$ is injected. The results show that despite an immediate activation, the initial number of non-active CAR T cells is not enough to clear the tumor which starts growing again after approximately 80 days.  
\begin{figure}[!ht]
    \centering    \includegraphics[width=0.8\columnwidth, trim={1cm 6cm 2cm 7cm},clip]{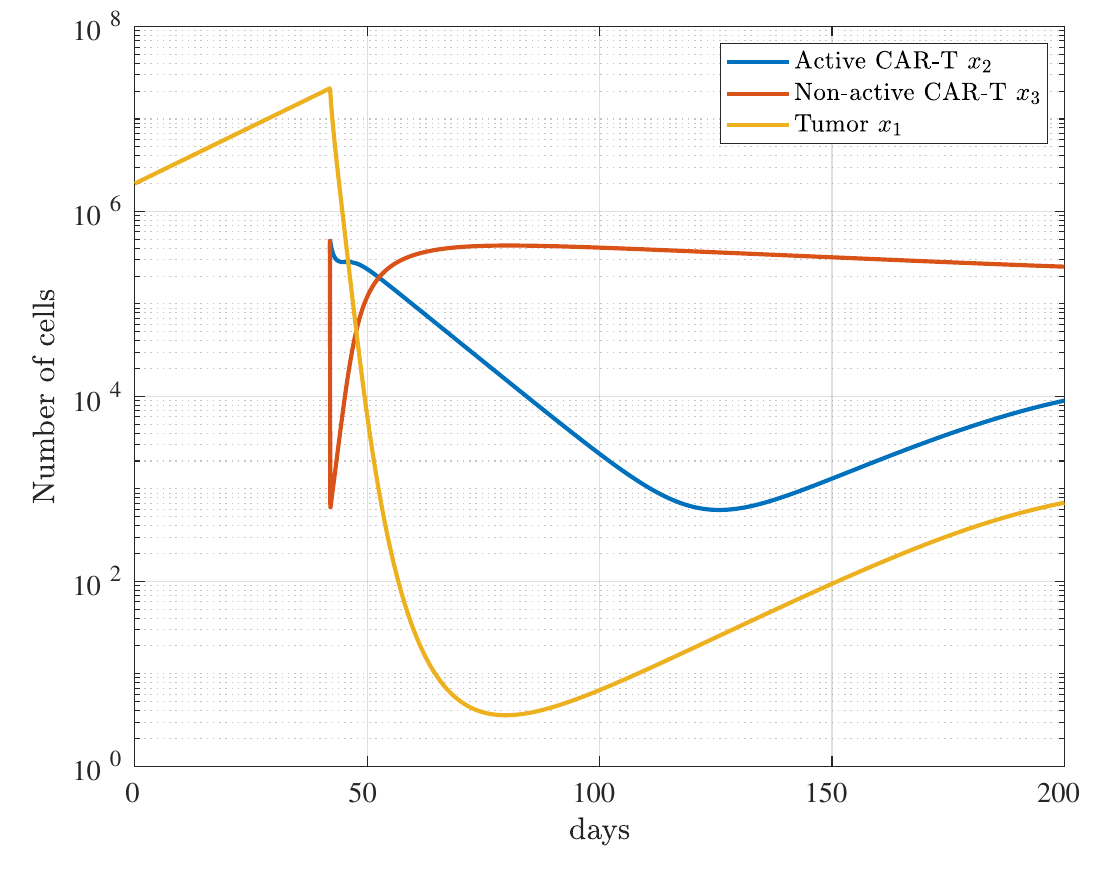}
    \caption{Model results starting with arbitrary initial conditions without the control input $\tau$.}
    \label{fig:no_control}
\end{figure}
The only equilibrium point with $x_1=0$ is denoted as $x_e^0$, and it is inherently unstable. This implies that relying solely on the initial conditions of non-active CAR T cells, represented by $x_3(0)$, is not a robust solution in the presence of uncertainties. Instead, by taking into account the constant input $\tau$ computed in \eqref{tau}, the original equilibrium point $x_e^0$ becomes a new equilibrium point, $x_n^e$, as defined in \eqref{new_eq}. As illustrated in Fig. \ref{fig:sim_bks}, we observe that the same initial conditions as those used in the simulation depicted in Fig. \ref{fig:no_control} can guide the system into the region of attraction centered around $x_1=0$. The introduction of $\tau$ effectively drives the tumor cell population towards asymptotic clearance, ensuring tumor elimination. Based on our simulations and the steady-state values $\bar x_2$ and $\bar x_3$, we can select a suitable value for the parameter $k=6$ \eqref{bound_k}. This choice results in $\xi=1.6946e+03$, demonstrating that \eqref{Lyapunov_f} is indeed a decreasing function.
\begin{figure}[!ht]
    \centering    \includegraphics[width=0.8\columnwidth, trim={1cm 6cm 2cm 7cm},clip]{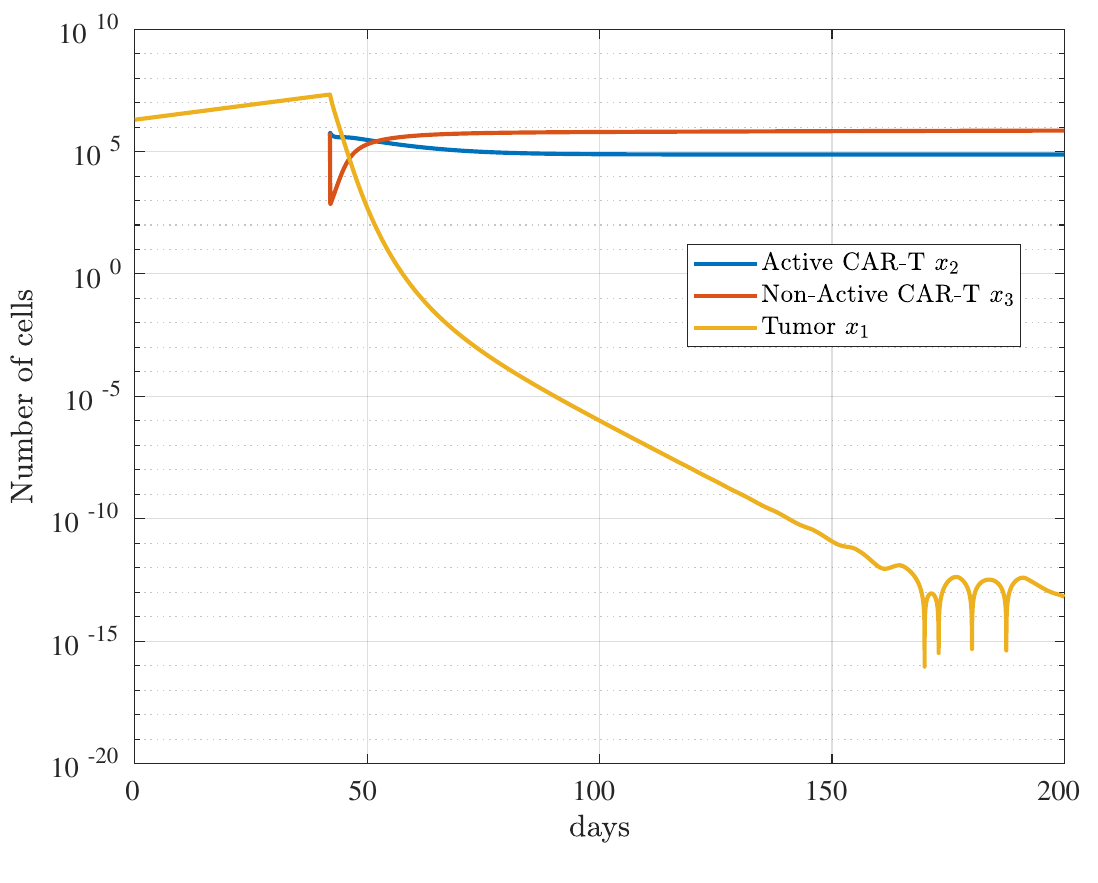}
    \caption{Model results with $\tau$ equal to \eqref{tau} with $a>1$.}
    \label{fig:sim_bks}
\end{figure}
It is important to emphasize that the level of tumor clearance achieved by the new scheme, as illustrated in Fig. \ref{fig:sim_bks_1}, is equivalent to that attained through uncontrolled CAR T cell activation, as shown in Fig. \ref{fig:sim_bks_1}. However, in the uncontrolled case (Fig. \ref{fig:sim_bks_1}), these cell populations exhibit indefinite growth, which can have detrimental consequences for the entire system, potentially leading to fatal toxicity in patients.
\begin{figure}[!ht]
    \centering    \includegraphics[width=0.8\columnwidth, trim={1cm 6cm 2cm 7cm},clip]{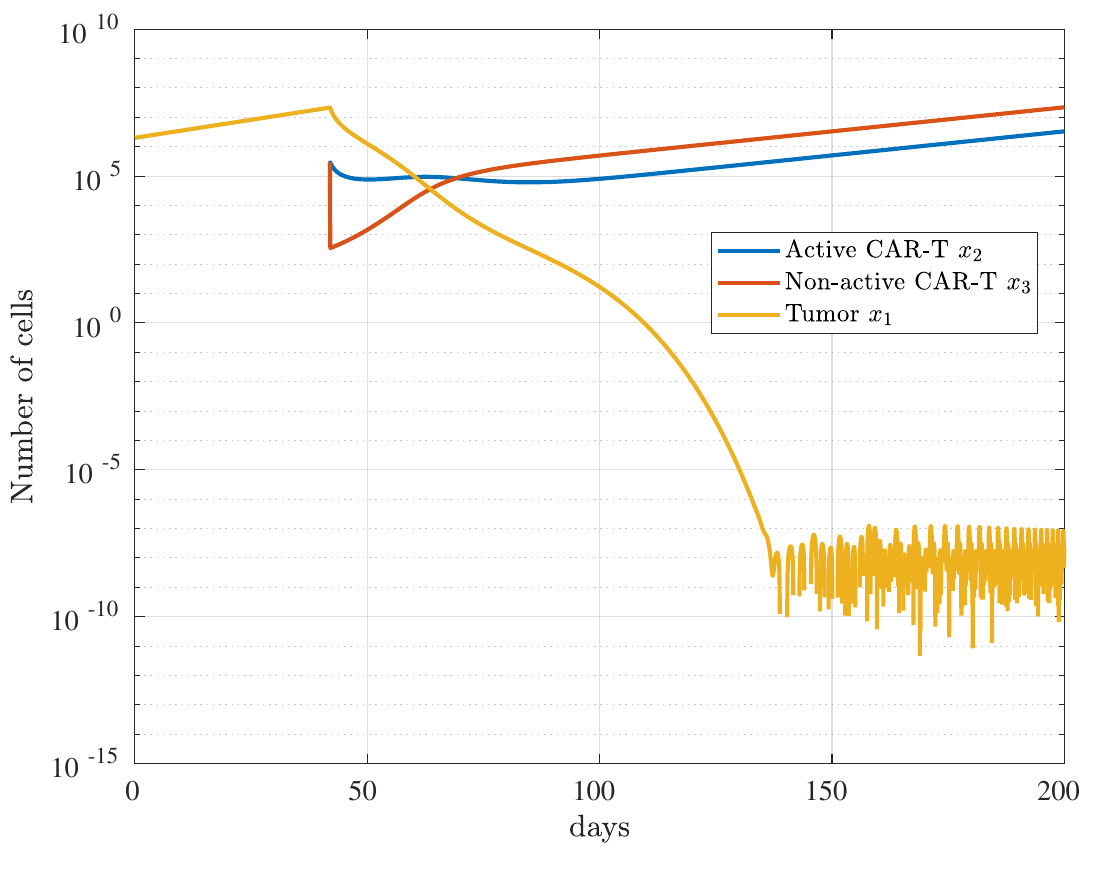}
    \caption{Model behavior with uncontrolled CAR T cell activation.}
    \label{fig:sim_bks_1}
\end{figure}}
\section{Conclusions}
Control theory holds significant promise in shaping the development of synthetic biological gene circuits and systems. In this context, our attention was directed towards a prominent player in synthetic biology therapy for cancer CAR T therapy. We harnessed control theory principles to analyze an established mathematical model, seeking novel avenues to enhance its therapeutic efficacy. Our findings demonstrated that by effectively aligning the initial conditions of non-active CAR T cells with a therapy designed to bolster CAR T cell activation, we can guarantee tumor clearance. The work suggests key therapeutic features with practical design potential. Such features include building in gene circuitry to mediate external activation and expansion of CAR T cells separate from tumor antigen-mediated activation, to accomplish feedforward CAR T cell activation, and to dampen CAR T cell activation. This work demonstrates the utility of applying control theory concepts to guide synthetic gene circuit design toward non-intuitive functions.

\section{Acknowledgement}
The author would like to thank Dr. Jason Lohmueller and Victor So from the Department of Surgery at the University of Pittsburgh, UPMC Hillman Cancer Center, for their valuable discussions and support for this work.
\bibliographystyle{IEEEtran}
\bibliography{references}
\end{document}